\documentclass[aps, prd, twocolumn, groupedaddress ]{revtex4-1}
\usepackage{amsmath,amssymb}
\usepackage{graphicx}
\usepackage{color}
\usepackage{slashed} 
\usepackage{hyperref}
\begin{document}
\newcommand{\nn}{\nonumber}
\def\d{{\mathrm{d}}}
\def\lint{\hbox{\Large $\displaystyle\int$}}   
\def\hint{\hbox{\huge $\displaystyle\int$}}  
\title{\bf\Large Physical observability of horizons}
\author{Matt Visser\,}
\email[]{matt.visser@msor.vuw.ac.nz}
\affiliation{ \mbox{School of Mathematics, Statistics, and Operations Research,}
Victoria University of Wellington; \\
PO Box 600, Wellington 6140, New Zealand.\\}
\date{28 July 2014; 5 August 2014; 29 August 2014; 15 November 2014; \LaTeX-ed \today}
\begin{abstract}
Event horizons are (generically) not physically observable. In contrast, \emph{apparent} horizons (and the closely related \emph{trapping} horizons) \emph{are} generically physically observable --- in the sense that they can be detected by observers working in  finite-size regions of spacetime.   Consequently \emph{event} horizons are inappropriate tools for defining \emph{astrophysical} black holes, or indeed for defining any notion of  \emph{evolving} black hole,  (evolving either due to accretion or Hawking radiation). The only situation in which an event horizon becomes physically observable is for the very highly idealized stationary or static black holes, when the event horizon is a Killing horizon which is degenerate with the apparent and trapping horizons; and then it is the physical observability of the apparent/trapping horizons that is fundamental --- the event horizon merely comes along for the ride.

\end{abstract}
\maketitle

\section{Introduction}

Some 40 years ago Stephen Hawking predicted  that black holes will emit radiation and slowly evaporate due to subtle quantum physics effects~\cite{hawking74a, hawking74b}. 
This prediction continues to generate heated debate, both from within the scientific community, and (sometimes) in the popular press. 
See for instance Hawking's recent opinion piece regarding the necessity of making careful physical distinctions between the mathematical concepts of \emph{event} horizon and \emph{apparent} horizon~\cite{horizons}: 
``The absence of event horizons means that there are no black holes --- in the sense of regimes from which light can't escape to infinity. 
There are, however, apparent horizons which persist for a period of time.'' 
(Hawking's opinion piece was then grossly misrepresented in the popular press,  by the simple expedient of mis-quoting a key phrase; and specifically by suppressing crucial subordinate clauses and sentences.)  
Related ideas along these lines have been mooted before, both by Hawking himself, 
(``... a true event horizon never forms, just an apparent horizon.''~\cite{dublin}), 
and (to one degree or another) by many other researchers~\cite{Roman, Ashtekar, Hayward05a,Hayward05b, micro-survey, Hossenfelder:2009,  Frolov:2014, Israel:2014, Bardeen:2014}.

Indeed it is well-known that there are \emph{very many} quite  \emph{different} types of horizon one can define. At a minimum: the event, apparent, trapping, isolated, dynamical, evolving, causal, Killing, non-Killing, universal, Rindler, particle, cosmological, and putative horizons, \emph{etcetera}. (See for instance~\cite{Booth:2005, Andersson:2005, Killing1, Killing2, Krasinski:2003, Helfer:2011}.) 

It is less well appreciated that the precise technical differences between these horizons makes a difference when one is worried about the subtleties of the ``black hole" evaporation process. 
These distinctions  even make a difference when precisely defining what a ``black hole'' is --- the usual definition in terms of an \emph{event} horizon is mathematically clean, leading to many lovely theorems~\cite{Hell}, but bears little to no resemblance to anything a physicist could actually measure.  (Somewhat related issues also afflict cosmology~\cite{lost, Binetruy:2014}.)

So why have \emph{event} horizons dominated so much of this discussion over the last 40 years?
Certainly \emph{event} horizons are known to exist in classical general relativity~\cite{Hell, lost, Binetruy:2014, MTW, Wald, Hartle, Poisson}, but they are extremely delicate \emph{teleological} constructions, somehow implicitly defined by a ``final cause'', requiring that nature inherently tends toward definite ends~\cite{telos}. 
Mathematically, one needs to know the entire history of the universe, all the way into the infinite future, and all the way down to any spacelike singularity, to decide whether or not an \emph{event} horizon exists right here and now. 
This makes \emph{event} horizons unsuitable for empirical testing in either laboratories or telescopes. 

In contrast, \emph{apparent} and \emph{trapping} horizons are defined using local (or at worst quasi-local) measurements, meaning that they are at least in principle suitable for  testing in \emph{finite-size laboratories or telescopes}. 

\section{Detectability of apparent/ trapping horizons} 

Consider for definiteness a dynamic spherically symmetric spacetime. Without any significant loss of generality we can write the metric as:
\begin{eqnarray}
\d s^2 &=& - \zeta(r,t)^2 \left\{1-{2m(r,t)\over r}\right\} \d t^2 + {\d r^2\over1-2m(r,t)/r} 
\nonumber\\
&&\qquad\qquad + r^2 (\d\theta^2+\sin^2\theta\;\d\phi^2).
\end{eqnarray}
Detecting event horizons requires knowledge of $m(r,t)$ throughout the entire spacetime, and in particular into the infinite future. Detecting apparent horizons, (and the closely related trapping horizons), requires one to evaluate the expansion
$\Theta \propto \{1-2m(r,t)/ r\}$
of outward pointing null geodesics, and boils down~\cite{evolving} to making a ``local'' measurement of the quantity $2m(r,t)/r$. 

To proceed, it is both useful and important to clarify the meaning of the much abused word ``local'' by carefully distinguishing the concept ``ultra-local" (measurements made at a single point) from  ``quasi-local'' (measurements made in a finite-size region of spacetime; that is, using a finite-size laboratory over a finite time interval). 
The equivalence principle, (more specifically the local flatness of spacetime), implies that no ultra-local measurement can (even in principle) \emph{ever} detect \emph{any} type of horizon. 
(Similarly, no ultra-local measurement can, even in principle, detect spacetime curvature.)

The situation is quite different for quasi-local measurements.
In particular, it is well-known that in any finite-size laboratory one can measure tidal effects, which are controlled by the Riemann tensor.  
Then in spherical symmetry, regardless of how one chooses to set up an orthonormal basis in the $r$--$t$ plane, in the $\theta$--$\phi$ plane one has:
\begin{equation}
R_{\hat\theta\hat\phi\hat\theta\hat\phi} = {2m(r,t)\over r^3}.
\end{equation}
The point is that $2m(r,t)/r^3$ is certainly measurable in a finite-size laboratory.  Furthermore, by measuring the extent to which the ``verticals'' converge on each other, a finite-size laboratory can also directly measure the radial coordinate $r$. Consequently in any finite-size laboratory the quantity $2m(r,t)/r$ is physically measurable. 

Detection of the presence (or absence) of an \emph{apparent} horizon is therefore a well-defined quasi-local observable.
Similar comments apply to the very closely related notion of \emph{trapping} horizon~\cite{Booth:2005, Andersson:2005, Poisson, evolving}.
For historical reasons, Hawking himself prefers to phrase his discussion in terms of \emph{apparent} horizons. However, formal developments over the last two decades or so suggest that a rephrasing in terms of \emph{trapping} horizons is potentially more fruitful. In particular, some non-symmetric foliations of spherically symmetric spacetimes can lead to rather ``unexpected'' apparent horizons~\cite{Wald:1991}, and the trapping horizons are better behaved in this regard. (See eg~\cite{Booth:2005, Andersson:2005}.)
In short, detection of the presence (or absence) of an \emph{apparent} or \emph{trapping} horizon is a well-defined quasi-local observable.

\section{Event horizons in flat spacetime} 

A well-known but often under-appreciated aspect of event horizons is that (contrary to common misperceptions) they are \emph{not} necessarily associated with strong gravitational fields; at least not in any simple local manner. 
In fact it is possible to arrange a situation in which a classical event horizon forms in a portion of spacetime that is a zero-curvature Riemann-flat segment of Minkowski spacetime. 

To do this merely consider a spherically symmetric shell of dust with a vacuum interior. (See eg~\cite{Poisson}, and figure 1.) 
By the Birkhoff theorem, the spacetime inside the shell is a segment of Minkowski space, and the spacetime outside the shell is a segment of Schwarzschild spacetime, (with some fixed constant mass parameter $m$). 
As the dust shell collapses, it will eventually cross its Schwarzschild radius and a black hole will form. Consider the resulting horizon. 
Then, as expected, to the future of the Schwarzschild-radius-crossing event the horizon is indeed an event horizon which is located at $r=2m$. 
However, there is also a portion of the event horizon that stretches backwards from the Schwarzschild-radius-crossing event to $r=0$, the centre of the spacetime. 
This portion of the event horizon is located in a segment of flat Minkowski space. 

Perhaps even worse, the very existence of this event horizon is predicated on both the eternal immutability of the resulting black hole, (infinitely far into the future), \emph{and} on assuming that the Schwarzschild solution continues to hold \emph{exactly}, all the way down to the central spacelike singularity.

(There is also an apparent horizon which first forms when the shell crosses its Schwarzschild radius, and then immediately splits, with one part of the apparent horizon remaining at $r=2m$ while the second part sits on top of the dust shell and follows it down to the centre at $r=0$).

\begin{figure}[!htbp]
\begin{center}
\includegraphics[width=4cm, height=5cm]{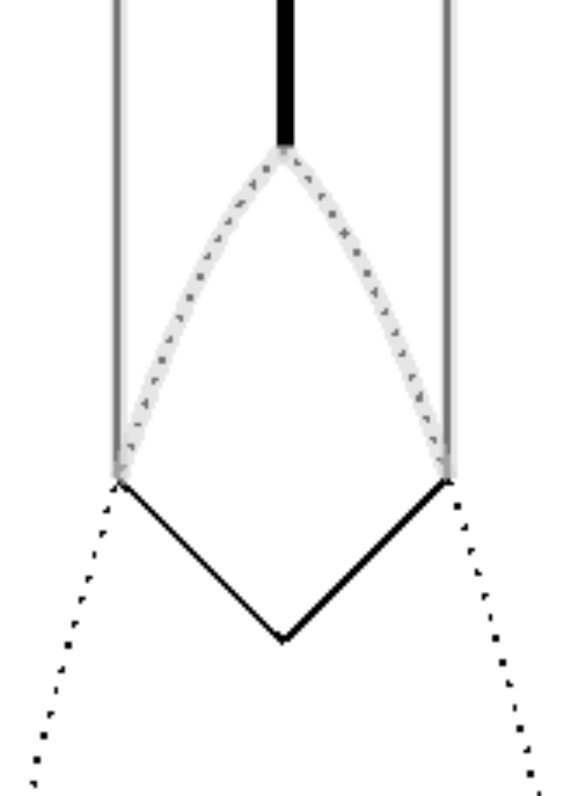}
\caption{The event horizon (thin solid line) reaches back into the flat Minkowski region. Part of the apparent horizon (grey line) follows the collapsing dust shell (dotted line) down to the singularity (thick black line).\vspace{-10pt} }
\label{F:1}
\end{center}
\end{figure}

\section{Non-detectability of event horizons} 

Consider a spherical laboratory of radius $R$ that collects data for some time duration $T$. 
Assume the equipment in the laboratory to have negligible mass, so that the gravitational self-field of the laboratory can be neglected. 
Surround the laboratory with a spherical dust shell; by the Birkhoff theorem all quasi-local experiments inside the laboratory will continue to detect a Minkowski spacetime. 

Now choose the dust shell such that the mass parameter of the segment of Schwarzschild spacetime outside the dust shell satisfies
\begin{equation}
2m > R + T.
\end{equation}
Let the dust shell go, drop it, arranging the timing so that the dust shell crosses its Schwarzschild radius just as the laboratory stops collecting data. 
Then the resulting event horizon reaches back to engulf the entire laboratory over the entire time interval that it was collecting data. 
There is simply no way that denizens of the laboratory could have detected the presence of the event horizon by any quasi-local means.

Related comments can be found (for instance) in work by Bengtsson and Senovilla~\cite{Bengtsson:2010, Bengtsson:2008}, and in the Living Review of Ashtekar and Krishnan~\cite{Ashtekar:2004}, and in Hayward's article from the year 2000~\cite{Hayward:2000}. It is perhaps a little disturbing to realise that the quite serious deficiencies and limitations exhibited by  \emph{event} horizons, while well appreciated within the general relativity community, are largely not understood or appreciated in the wider physics community.

\section{Destroying horizons}

It is again well-known, if not widely appreciated, that from the point of view of QFT the Hawking flux can be viewed as a \emph{negative} flux \emph{into} the black hole~\cite{B&D}.   
This suggests a simple toy model, a gedanken-experiment,  that can be used to capture \emph{some} of the key features of Hawking evaporation. 
Let us  now surround the black hole we have just considered with a second shell of \emph{negative} mass dust, carefully tuned to make the total mass of the system zero. 
(So, again by the Birkhoff theorem, inside the inner shell the spacetime is a segment of Minkowski spacetime, between the two shells it is a segment of Schwarzschild spacetime with some fixed mass parameter $m$, and outside the outer shell it is again a segment of Minkowski spacetime.) 
As the negative-mass outer shell drops, it will eventually cross $r=2m$, in the process  destroying the horizon that had (previously and temporarily) formed due to the inner shell crossing its Schwarzschild radius. See figure 2.
(See also the related discussion in reference~\cite{Hayward05a}, and in the book by Frolov and Novikov~\cite{Frolov-Novikov}.)

But what sort of horizon is/was this? 
Certainly there is an apparent horizon in this spacetime. 
As previously, the apparent horizon first forms when the inner shell crosses its Schwarzschild radius,  immediately splitting, with one part of the apparent horizon remaining at $r=2m$ while the second part rides the inner shell down to the centre at $r=0$ (where a curvature singularity forms). 
Later on, once the outer shell drops down to $r=2m$, the apparent horizon, (suitably defined), follows  the outer shell down to the centre at $r=0$. (Where the curvature singularity then unforms.) 
But where is the event horizon? 

Go to the event where the outer shell finally reaches the centre, $r=0$, and then in the causal past of that event use the Schwarzschild geometry to backtrack the ``outgoing'' (ie, less ingoing) null curve into the past, until it eventually crosses the inner shell. 
This null curve will always strictly satisfy $r<2m$, and so will be strictly inside (though asymptotically approaching) the apparent horizon. 
Once this backtracked null curve crosses the inner shell, from that event backtrack (now in the Minkowski geometry) the outgoing radial null curve until one reaches the centre $r=0$.  
This segment of the event horizon always lies strictly outside the relevant segment of the apparent horizon. 

\begin{figure}[!htbp]
\begin{center}
\includegraphics[width=4cm, height=5cm]{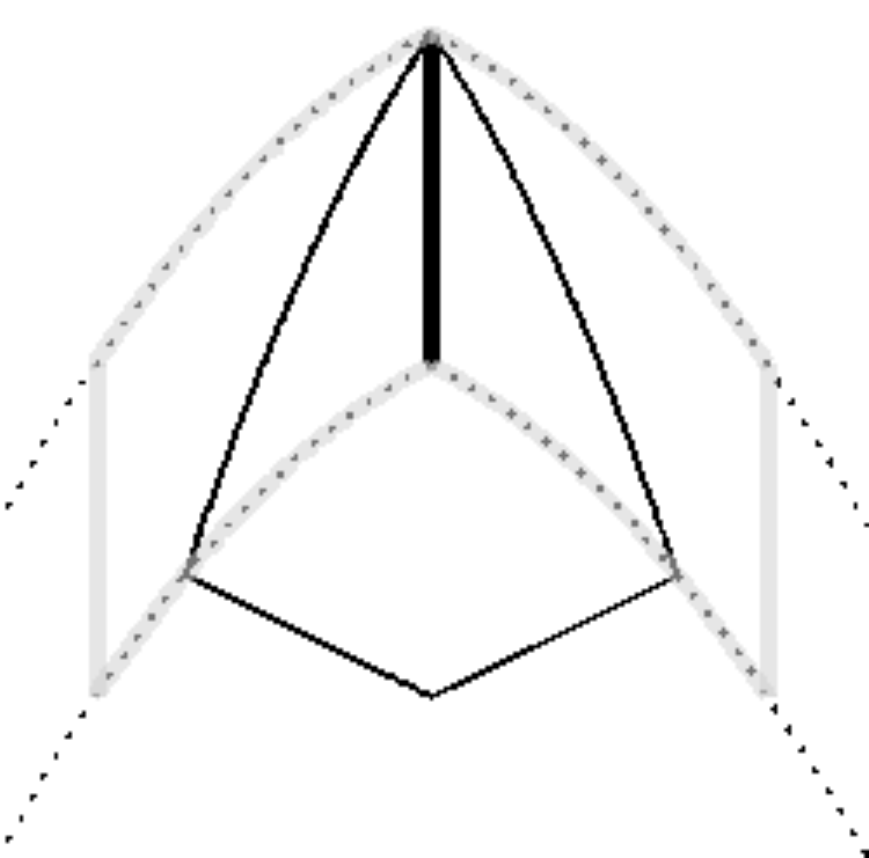}
\caption{Part of the event horizon (thin black line) reaches back into the flat Minkowski region, and the other part reaches forward to the point where the singularity (thick black line) is destroyed. Part of the apparent horizon (grey line) follows the two collapsing dust shells (dotted lines) down to the singularity, the remaining portion lies at $r=2m$ between the two shells. The event horizon and apparent horizon occur at qualitatively different locations. \vspace{-15pt} }
\label{F:2}
\end{center}
\end{figure}

So in this toy model there is still an event horizon, but it is not where one might have thought it should be, and in particular does not extend to the infinite future. 
Part of the event horizon lies in a Riemann-flat section of spacetime, and the part of the event horizon that lies in the Schwarzschild segment of the spacetime does not occur at $r=2m$. 
One cannot even begin to calculate the location of the event horizon until one has final definitive information that the spacetime has stabilized into its final form. 
One could always posit yet a third dust shell, waiting in the wings, ready to engulf the entire region of interest with an unexpected event horizon. 
In counterpoint, there are a number of models in which the event horizon never forms in the first place.

\section{Excluding horizons completely} 

The distinction between the apparent and event horizons can be so non-perturbatively extreme as to completely exclude event horizons. 

In particular this is the logic underlying the physics of the Bergmann--Roman~\cite{Roman}, the Hayward~\cite{Hayward05a, Hayward05b}, and the more recent Frolov~\cite{Frolov:2014} scenarios, in all of which the apparent horizon is taken to be a topologically toroidal surface, $S^1\times S^2$, which does not intersect the always time-like $r=0$ world line. 
Some other models, such as that of Ashtekar--Bojowald~\cite{Ashtekar}, are based on apparent horizons that reach down to $r=0$, (and so are topologically $\mathbb{R}\times S^2$).  
Something more or less along these lines is also the underlying gestalt for Hawking's recent article~\cite{horizons}, and the models of Israel~\cite{Israel:2014} and  Bardeen~\cite{Bardeen:2014}.

\section{Killing horizons} 

Of course in static spacetimes and stationary spacetimes the \emph{event}, \emph{apparent}, \emph{trapping}, and \emph{Killing} horizons happen to coincide. 
Thereby, physical detectability of  \emph{apparent}, \emph{trapping}, and \emph{Killing} horizons  implicitly implies  the physical detectability of the \emph{event} horizon; but only for this restricted class of spacetimes. (See for example the discussion in reference~\cite{Poisson}.)
In non-stationary spacetimes \emph{Killing} horizons no longer exist, and the \emph{apparent} and \emph{trapping} horizons need not, and typically will not, coincide with the \emph{event} horizon, even if one might exist. 
Nor, as we have seen above,  need any event horizon even be ``perturbatively close'' to any apparent/trapping horizon. 

\section{ADAFs} 

The empirical astronomical observation of ``advection dominated accretion flows''~\cite{adaf, Narayan, Narayan:2008, Narayan:1998} is often interpreted as evidence for the existence of \emph{event} horizons ---  but ADAFs cannot distinguish between event horizons and apparent/trapping horizons. 
The key point is that one  observes the \emph{lack} of bremsstrahlung as infalling matter coalesces with the putative black hole --- indicating that the surface is not some solid/liquid/gas interface, but is instead acting as some sort of ``one way membrane'' (at least for the temporal duration of the observations). 
The ADAF observations are thus suggestive of the presence of \emph{some sort} of horizon, but cannot be used to specifically argue for the presence of \emph{event} horizons.

\section{Discussion}

Perhaps surprisingly to non-relativists, there are \emph{very many quite different} types of horizon one can define. 
(At the very least, one can define the event, apparent, trapping, isolated, dynamical, evolving, causal, Killing, non-Killing, universal, Rindler, particle, and putative horizons, \emph{etcetera}...~\cite{Booth:2005,  Andersson:2005, Killing1, Killing2, Hell, Krasinski:2003, Helfer:2011, lost, Binetruy:2014, MTW, Wald, Hartle, Poisson, evolving}.) 

Furthermore, the precise technical differences between these horizons sometimes can (and often does) make a physical difference in physically relevant and mathematically well-defined situations. 
Additionally, again perhaps surprisingly to non-relativists, \emph{event} horizons, (despite their undoubted mathematical virtues in allowing one to prove quite powerful general theorems~\cite{Hell} in classical general relativity), are in any dynamical context not physically detectable by any finite-size laboratory or telescope, severely curtailing their empirical usefulness.

So let us focus on issues  that can at least in principle be testable in a finite-size laboratory. 
The point is that while event horizons are certainly properties of classical general relativity (Schwarzschild spacetime, Kerr spacetime, and their evolving generalizations, \emph{etcetera}), there is little to no evidence that \emph{event} horizons survive the introduction of even semi-classical quantum effects (let alone full quantum effects).  
\emph{Event} horizons are extremely delicate and depend on global geometry --- including what will happen in the infinite future. 
In contrast  \emph{apparent} horizons or \emph{trapping} horizons seem much more robust in this regard --- they depend on local or quasi-local physics and are much more difficult to destroy. 
(Some authors prefer to use the more general term ``quasi-local horizons''.)

In particular, in somewhat different contexts, it has already been demonstrated  that the presence of an \emph{event} horizon~\cite{essential}, 
or indeed \emph{any} sort of horizon~\cite{minimal1, minimal2, trapped} is not essential for Hawking radiation and ``black hole evaporation'' to occur,  and for that matter, in some non-GR analogue spacetime contexts one can even dispense with black hole entropy 
(Bekenstein entropy)~\cite{analogue, acoustic, no-entropy}. 

The tangle of issues related (both directly and indirectly) to these observations has significant implications for all of black hole thermodynamics, and in particular is central to the understanding of the endpoint of the Hawking evaporation process~\cite{firewall, firewall2, Braustein:2009, no-firewall, no-firewall2, no-firewall3, no-firewall4, no-firewall5, no-firewall6, no-firewall7, no-firewall8, no-firewall9}. 

\acknowledgments

This research was supported by the Marsden Fund, and by a James Cook fellowship, both administered by the Royal Society of New Zealand.



\end{document}